\def\nocolor{nocolor}
\ifx\scipostversion\undefined 
  \documentclass[aps,pra,twocolumn,superscriptaddress,floatfix]{revtex4-1}
\else
  \documentclass[submission, Phys]{SciPost}
\fi

\ifx\scipostversion\undefined  
\fi   

\usepackage{graphicx}
\usepackage{amsmath}
\usepackage{bbold}
\usepackage[T1]{fontenc}
\usepackage{hyperref} 

\ifx\nocolor\undefined 
   \usepackage{xcolor}
   
   \newcommand{\tcolr}[1]{\textcolor{red}{#1}}
\else
   
   \newcommand{\tcolr}[1]{{#1}}
\fi

\newcommand*{\dt}{\mathrm{d}t}
\newcommand*{\dth}{\frac{\mathrm{d}t}{2}}

\newcommand*{\cA}{\mathcal{A}}

\newcommand{\beq}{\begin{equation}}
\newcommand{\eeq}{\end{equation}}
\newcommand{\beqs}{\begin{equation*}}
\newcommand{\eeqs}{\end{equation*}}
\newcommand{\beqa}{\begin{eqnarray}}
\newcommand{\eeqa}{\end{eqnarray}}
\newcommand{\beqas}{\begin{eqnarray*}}
\newcommand{\eeqas}{\end{eqnarray*}}
\def\bals#1\eals{\begin{align*}#1\end{align*}}
\def\bal#1\eal{\begin{align}#1\end{align}}

\newcommand{\bcent}{\begin{center}}
\newcommand{\ecent}{\end{center}}

\newcommand{\bitem}{\begin{itemize}}
\newcommand{\eitem}{\end{itemize}}

\newcommand{\phe}{\phantom{e}}
\newcommand{\phs}{\phantom{i}}
\newcommand{\pht}{\phantom{,}}
\newcommand{\phn}{\phantom{n}}
\newcommand{\phm}{\phantom{m}}

\newcommand{\ioh}{\frac{i}{\hbar}}

\newcommand{\one}{\mathbb{1}}

\makeatletter
\newcommand*\bt{\mathpalette\bt@{.7}}
\newcommand*\bt@[2]{\mathbin{\vcenter{\hbox{\scalebox{#2}{$\m@th#1\bullet$}}}}}
\makeatother

\makeatletter
\newcommand*\ct{\mathpalette\ct@{.7}}
\newcommand*\ct@[2]{\mathbin{\vcenter{\hbox{\scalebox{#2}{$\m@th#1\circ$}}}}}
\makeatother

\newcommand{\Arm}{\mathrm{A}}
\newcommand{\Srm}{\mathrm{S}}
\newcommand{\Hrm}{\mathrm{H}}

\newcommand*{\covdev}{\text{\dh}}

\begin{document}

\title{Numerical integration of quantum time evolution in a curved manifold}
             
\author{Jessica F. K. Halliday}
\affiliation{Theory of Condensed Matter,
             Cavendish Laboratory, University of Cambridge, 
             J. J. Thomson Ave, Cambridge CB3 0HE, United Kingdom}             

\author{Emilio Artacho}
\affiliation{Theory of Condensed Matter,
             Cavendish Laboratory, University of Cambridge, 
             J. J. Thomson Ave, Cambridge CB3 0HE, United Kingdom}
\affiliation{CIC Nanogune BRTA and DIPC, Tolosa Hiribidea 76, 
             20018 San Sebastian, Spain}
\affiliation{Ikerbasque, Basque Foundation for Science, 48011 Bilbao, Spain}

\date{\today}

\begin{abstract}
  The numerical integration of the Schr\"odinger equation by discretization of time
is explored for the curved manifolds arising from finite representations based on
evolving basis states.
  In particular, the unitarity of the evolution is assessed, in the sense of the 
conservation of mutual scalar products in a set of evolving states, and with them
the conservation of orthonormality and particle number.
  Although the adequately represented equation is known to give rise to unitary
evolution in spite of curvature, discretized integrators easily break 
that conservation, thereby deteriorating their stability.
  The Crank Nicolson algorithm, which offers unitary evolution in Euclidian
spaces independent of time-step size $\mathrm{d}t$, can be generalised to curved
manifolds in different ways.
  Here we compare a previously proposed algorithm that is unitary by
construction, albeit integrating the wrong equation, with a faithful
generalisation of the algorithm, which is, however, not strictly unitary
for finite $\mathrm{d}t$.
\end{abstract}

\pacs{}

\maketitle

\section{Introduction}

  Algorithms for the efficient numerical integration of the time-dependent 
Schr\"odinger equation in discretised real time for finite representations 
have been discussed at some length in the literature (see e.g. 
Refs~\cite{Kosloff1991, Castro2004, Correa, Varga2017}), normally 
in the context of the single-particle states in time-dependent Hartree-Fock 
or time-dependent density-functional theory (TD-DFT) \cite{RungeGross}.
  The consideration of evolving basis sets complicates matters, and
there is less knowledge accumulated on good integrators for them and for
arbitrarily quick basis evolution \cite{Sankey,Andermatt2016, Kaxiras2015,
Artacho2017,Hekele2020}.
    Evolving basis sets are  routinely encountered in electronic structure 
 calculations for which
atom-centered basis functions are used, and where atoms move, that is,
any first-principles dynamical calculation method in quantum chemistry or 
condensed matter and materials physics using atomic orbitals as
basis sets. 
  There are many such software packages that are widely used in
either or both communities.
  For a brief review and links to codes used in 
quantum chemistry see, e.g.,  Ref.~\cite{Handbook2017}; for 
methods and programs using atomic orbitals in condensed matter 
see, e.g., Refs.~\cite{Garcia2020, Kuhne2020, Oliveira2020, 
Dovesi2020, QuantumATK2019, Blum2009, Delley2000}.
  
  The equation for the evolution of quantum states for a moving basis is 
easily obtained.
  For a basis set $\{ |e_{\mu}, t\rangle, \, \mu = 1 \dots \mathcal{N} \}$, 
$H|\psi_n\rangle = i\hbar\partial_t |\psi_n\rangle$ straightforwardly becomes
\beq
\label{eq:tdse}
\sum_{\nu}^{\mathcal{N}} H_{\mu\nu} \psi^{\nu}_{\phe n} = i \hbar 
\sum_{\nu}^{\mathcal{N}} ( S_{\mu\nu} \partial_t \psi^{\nu}_{\phe n} + 
D_{\mu\nu t} \psi^{\nu}_{\phe n} ) \, ,
\eeq
with $H_{\mu\nu} = \langle e_{\mu} | H | e_{\nu}\rangle$, 
$S_{\mu\nu} = \langle e_{\mu} | e_{\nu}\rangle$, 
$D_{\mu\nu t} = \langle e_{\mu} | \partial_t | e_{\nu}\rangle$, and
$\psi^{\nu}_{\phe n}$ the coefficients in the expansion
\beq
\label{eq:expand}
|\psi_n\rangle = \sum_{\nu}^{\mathcal{N}} |e_{\nu}\rangle \psi^{\nu}_{\phe n} \, .
\eeq
   
   It is known that the evolution of a set of states following this equation
 is unitary in the sense that it 
 preserves their mutual scalar products (see e.g. \cite{Todorov2001}).
 Therefore, if the evolving states are, for instance, the occupied Kohn-Sham
 states in TD-DFT evolution, they preserve their orthonormality, and the 
 number of particles is conserved.
 
  It is not obvious, however, how to guarantee such unitarity for
approximate algorithms based on time discretization.
  Notice that the $D_{\mu\nu t}$ matrix does not need to be
anti-hermitian if the evolution of basis vectors $|e_{\mu}\rangle$
and $|e_{\nu}\rangle$ is arbitrary (think e.g. of one of them not evolving
while the other does), \tcolr{and therefore the usual thinking in terms of 
unitary matrices from the exponential of hermitian matrices does
not apply, at least directly}. 
  In this work we focus on how the unitarity of the time evolution
is affected by the discretization of time for numerical integration. 

  The semiclassical description of atomic collisions has made use of
travelling orbitals \cite{Errea1979}, defined as
\beqs
\langle \mathbf{r} | e_{\mu},t\rangle = e^{i m_e \mathbf{v}\cdot\mathbf{r}/\hbar}
f_{\mu} (\mathbf{r}-\mathbf{v} t) \, ,
\eeqs
where $f_{\mu} (\mathbf{r}) = R(r) Y_l^m(\theta,\phi)$, 
is a time-independent atomic-like orbital, and $\mathbf{v}$ is the velocity
at which it is travelling, normally attached to a nucleus.
  They are very well adapted to the situation in which the electrons
themselves travel with the atoms and, therefore, with the basis 
states, which is frequently the case in atomic collisions \cite{Errea1979}.
  They are not so useful beyond that realm, as in e.g. atoms
moving in metals, where the electrons are pushed around by a
moving nucleus, but do not necessarily accompany it.
  
  A more general procedure based on a L\"owdin \tcolr{orthonormalization} 
was proposed \tcolr{by Tomfohr and Sankey (TS hernceforth)} in 
Ref.~\cite{Sankey}, which was built strictly to preserve unitarity of 
evolution for finite time steps for any evolving basis.
  It has been quite successfully used for projectiles traversing solids at 
non-adiabatic but relatively low velocity \cite{Zeb2012, Zeb2013, Ullah2015, 
Halliday2019, Gu2020}. 
  The method was further discussed in Ref.~\cite{Artacho2017}, where it was 
shown to affect the equation being integrated, with the potential to
lead to inaccurate propagation for high velocity, even in the converged 
low time-step limit. 
%
  
  Finally, a different way of approaching the integration of Eq.~\eqref{eq:tdse}
is by rewriting it as 
\tcolr{
\beqs
\sum_{\nu}^{\mathcal{N}} (H_{\mu\nu} -i\hbar D_{\mu\nu t}) \psi^{\nu}_{\phe n} = i\hbar
\sum_{\nu}^{\mathcal{N}} S_{\mu\nu} \partial_t \psi^{\nu}_{\phe n}  \, ,
\eeqs
}and taking $H_{\mu\nu} - i \hbar D_{\mu\nu t}$ as
a modified Hamiltonian matrix (see e.g. Ref.~\cite{Andermatt2016}).
  The behavior for finite time step of this pragmatic approach is not
easy to discern from general considerations, since the $D_{\mu\nu t}$ matrix 
is not expected to be anti-hermitian, as mentioned above.
  It does work reasonably well, however \cite{Andermatt2016}.
  Here we explore it further, both formally, and by explicitly comparing its
time-step convergence with the very stable \tcolr{TS integrator \cite{Sankey}}, 
while the accuracy of the latter is
further scrutinized.

  For a better understanding of  the evolution we use in the following the
recent geometric interpretation \cite{Artacho2017} of Eq.~\eqref{eq:tdse}.
  The same paper proposed a strictly unitary integrator for an evolving basis, 
as long as the spanned Hilbert space $\Omega$ were invariant at all times,
which is approximately the case for a well converged basis set.
  Here we explore the situation for an evolving Hilbert space $\Omega(t)$,
defining a curved fibre bundle \cite{Artacho2017}, 
for the general situation in which its curvature cannot be neglected.

  \tcolr{Ref.~\cite{Artacho2017} identifies the $D_{\mu\nu t}$ matrix as a 
connection in differential geometry.
  The integration procedure of Ref.~\cite{Andermatt2016} can then be 
interpreted in this context as using the connection as a gauge potential 
in the Hamiltonian~\cite{Artacho2017}. 
  We will hence refer to this procedure as the gauge-potential (GP) integrator.}


\section{Unitarity in the equations}

  Following Ref.~\cite{Artacho2017}, the expansion in Eq.~\eqref{eq:expand} of any
quantum state in a non-orthogonal and evolving basis set, 
$\{ |e_{\mu}, t\rangle, \, \mu = 1 \dots \mathcal{N} \} $, defines an
evolving $\mathcal{N}$-dimensional Hilbert space $\Omega(t)$, which, in turn,
defines a $(\mathcal{N}+1)$-dimensional fibre bundle $\Xi$, which 
represents a non-Euclidian manifold.
  In its natural representation \cite{Artacho1991}, and summing over repeated indices,
Eq.~\eqref{eq:tdse} becomes \cite{Artacho2017}
\beq
\label{eq:tdsenat}
H^{\mu}_{\phantom{e}\nu} \psi^{\nu}_{\phs n}=i\hbar \, \covdev_t 
\psi^{\mu}_{\phs n} \; ,
\eeq
with 
\beqs
\psi^{\mu}_{\phs n} = \langle e^{\mu} | \psi_n\rangle \quad \mathrm{and}  \quad
H^{\mu}_{\phe\nu} = \langle e^{\mu} | H | e_{\nu} \rangle \, .
\eeqs
  The set $\{ |e^{\mu}, t\rangle, \, \mu = 1 \dots \mathcal{N} \}$ is the
dual basis of $\{|e_{\mu},t\rangle\}$, also a basis of $\Omega(t)$, 
satisfying 
$\langle e^{\mu},t |e_{\nu},t \rangle = \delta^{\mu}_{\phs\nu}, \, \forall \mu,\nu$
at any time $t$.
$\covdev_t$ represents the covariant time derivative \cite{Artacho2017} defined as 
\beqs
\covdev_t  \psi^{\mu}_{\phs n} = \partial_t \psi^{\mu}_{\phs n} 
+ D^{\mu}_{\phe\nu t} \psi^{\nu}_{\phs n} \; , 
\eeqs
where $D^{\mu}_{\phe\nu t}  =  \langle e^{\mu} | \partial_t e_{\nu} \rangle$ 
gives the connection in the manifold (note the convention in the order of indices).

  Similarly, a bra evolves following  
\beqs
\langle \psi_m | H =  -i\hbar \partial_t \langle \psi_m | \, ,
\eeqs  
where we have made use of the hermiticity of the Hamiltonian operator.
 It is represented by \cite{Artacho2017}
\beq
\label{eq:tdsebnat}
 \psi_{m\nu} H^{\nu}_{\phantom{e}\mu} = -i\hbar \, \covdev_t
 \psi_{m\mu} = -i\hbar (\partial_t  \psi_{m\mu}
 +  \psi_{m\nu} D^{\nu}_{\phantom{e}t \mu} ) \; ,
\eeq
with $\psi_{m\nu}=\langle \psi_m | e_{\nu}\rangle$, and
\begin{equation}
\label{eq:signchange}
D^{\nu}_{\phantom{e}t \mu} = \langle \partial_t e^{\nu} | e_{\mu} \rangle 
= - D^{\nu}_{\phantom{e}\mu t} \; ,
\end{equation}
the latter equality being due to 
$\partial_t \langle e^{\nu} | e_{\mu}\rangle =0$.
\tcolr{Eq.~\eqref{eq:signchange} is the key for the unitarity of 
the propagation, replacing the conventional expectation of
hermiticity of $i D_{\mu\nu t}$}.

\subsection{Conservation of scalar products}

  We start by showing the expected \cite{Todorov2001} unitarity of the 
evolution in the manifold, defined here as conservation of scalar products 
\beq
\label{eq:conserv-prod}
\partial_t \langle\psi_n|\psi_m\rangle = 0 \, , \quad \forall \, m,n
\eeq
among the propagating 
states $\{|\psi_m\rangle, m=1\dots N_e\}$ at any time.
  The evolution of the coefficients for the ket and bra then is determined by
Eqs.~\eqref{eq:tdsenat} and \eqref{eq:tdsebnat}, as
\tcolr{
\beqa
\label{eq:evoldiff}
&\partial_t \psi^{\mu}_{\phantom{e}m} = - (\frac{i}{\hbar} \, H^{\mu}_{\phantom{e}\nu}
+ D^{\mu}_{\phantom{e}\nu t} ) \, \psi^{\nu}_{\phantom{e}m} \\
\label{eq:evoldiffbra}
&\partial_t \psi_{m\mu} = \psi_{m\nu} 
(\frac{i}{\hbar} \, H^{\nu}_{\phantom{e}\mu}
+ D^{\nu}_{\phantom{e}\mu t} )    \; .
\eeqa
}It is easy to check that the scalar products 
$\langle \psi_m | \psi_n\rangle = \psi_{m\mu} \psi^{\mu}_{\phe n}$
are preserved in time:
\tcolr{
\bals
& \partial_t \langle \psi_m | \psi_n \rangle 
= \partial_t (\psi_{m\mu} \, \psi^{\mu}_{\phe n})
=(\partial_t \psi_{m\mu}) \psi^{\mu}_{\phe n} +
\psi_{m\mu} (\partial_t \psi^{\mu}_{\phe n}) \\ 
&= \psi_{m\nu} (\ioh H^{\nu}_{\phs\mu} + D^{\nu}_{\phs\mu t}) \psi^{\mu}_{\phs n} 
- \psi_{m\mu} (\ioh H^{\mu}_{\phs\nu} + D^{\mu}_{\phs\nu t}) \psi^{\nu}_{\phs n}=0 \, .
\eals
}
%

   In addition to the Hamiltonian operator being hermitian, the unitarity of the 
 propagation is therefore direct consequence of Eq. \eqref{eq:signchange}.
   The natural representation does not recover the usual self-adjoint matrix shape,
 but offers quite transparent relations and derivations.
   If the dealings above seem a bit of a sleight of hand, Appendix~\ref{app:unitary-matrix}
reproduces the result in the matrix representation.


\section{Finite time step $\dt$}
\label{sec:constHD}

  After time discretisation for numerical integration, we are interested
in propagating the set of coefficients 
\begin{equation*}
\psi^{\mu}_{\phantom{e}m}(t) \rightarrow  
\psi^{\mu}_{\phantom{e}m}(t+\mathrm{d}t) 
\end{equation*} 
for finite d$t$, trying to maximise both the quality and the stability of 
whatever the algorithm we use.
  Preservation of the orthonormality of the propagating states
is key for that purpose.
  Here we will explore the behaviour of the Crank Nicolson algorithm
in the $\Xi$ curved manifold.

  For the finite time-step d$t$ we will neglect henceforth the 
time evolution of both the hamiltonian and the connection between
$t$ and $t+\dt$.
  This is compatible with various integration algorithms
such as extrapolating the Hamiltonian to $t+\dt/2$.
  Other algorithms, such as the self-consistent Crank Nicolson \cite{Kaxiras2008}
or self-consistent predictor-corrector schemes \cite{Bao2015}
may need further consideration.

\subsection{Unitary propagation for static basis}

  Let us start with a reminder of unitary propagation when
the basis set may be non-orthogonal but does not evolve,
and, consequently, the state manifold is a regular Hilbert space $\Omega$.
  It means that $D^{\mu}_{\phantom{e}\nu t}=
D^{\mu}_{\phantom{e}t\nu}=0$, $S_{\mu\nu}$ and 
$S^{\mu\nu}$ are constant, and the solutions for 
Eqs.~\eqref{eq:evoldiff} and \eqref{eq:evoldiffbra} are
\beqa
\label{eq:ketbraevol0}
\psi^{\mu}_{\phe m} (t+\dt) &= e^{- \dt \ioh
H^{\mu}_{\phe\nu}} \, \psi^{\nu}_{\phe m} (t) \nonumber \\ 
\psi_{m\mu} (t+\dt) &= \psi_{m\nu} (t) \,
e^{ \dt \ioh H^{\nu}_{\phe\mu}} \; .
\eeqa

   Scalar products among $|\psi_m\rangle$'s are preserved  
as the states evolve, 
\tcolr{\bal
\label{eq:demo-static}
\langle \psi_m(t+\dt) &| \psi_n (t+\dt) \rangle = \nonumber \\
&=\psi_{m\mu} (t+\dt)  \psi^{\mu}_{\phe n} (t+\dt) \nonumber \\
&= \psi_{m\nu} (t) \, e^{ \dt \ioh H^{\nu}_{\phe\mu}}  \, e^{- \dt \ioh
H^{\mu}_{\phe\sigma}} \, \psi^{\sigma}_{\phe n} (t) \nonumber \\
&=\psi_{m\mu} (t)  \psi^{\mu}_{\phe n} (t) 
= \langle \psi_m(t) | \psi_n (t) \rangle 
\eal
}as expected.
  Appendices~\ref{app:static-exp} and \ref{app:static-bra}
show the same in the matrix representation, and that the 
evolved bra in Eq.~\eqref{eq:ketbraevol0} remains the
bra of the evolved ket, respectively.

\subsubsection{Crank-Nicolson for a static basis} 
The approximate evolution given by 
\beq
\label{eq:ket-cn-static}
\psi^{\ct}_{\phe m} (t+\dt) = 
\left [ \one^{\ct}_{\phs\bt} + \ioh \dth H^{\ct}_{\phs\bt} \right ]^{-1}
\left ( \one^{\bt}_{\phs\bt} - \ioh \dth H^{\bt}_{\phs\bt} \right )
\psi^{\bt}_{\phs m} (t)
\eeq
is the direct generalisation (in the natural representation) of the 
usual Crank-Nicolson evolution step for orthonormal bases
\cite{Artacho2017}.
  We have abstracted the notation, with the circles standing for indices, 
to be contracted with contiguous ones if full, and always up with down.
  Eq.~\eqref{eq:ket-cn-static} can also be recast in the matrix 
representation as follows
\bals
&\psi^{\ct}_{m} (t+\dt) = \\
&= \left [ \one^{\ct}_{\phs\bt} + \ioh \dth S^{\ct\bt} H_{\bt\bt} \right ]^{-1} \! \! 
\left ( \one^{\bt}_{\phs\bt} - \ioh \dth S^{\bt\bt} H_{\bt\bt} \right )
\psi^{\bt}_{\phs m} (t) \\
&=\left [ \one^{\ct}_{\phs\bt} + \ioh \dth S^{\ct\bt} H_{\bt\bt} \right ]^{-1}
\! \! \! \! S^{\bt\bt} S_{\bt\bt}
\left ( \one^{\bt}_{\phs\bt} - \ioh \dth S^{\bt\bt} H_{\bt\bt} \right )
\psi^{\bt}_{\phs m} (t) \\
&=\left [ S_{\ct\bt} + \ioh \dth H_{\bt\bt} \right ]^{-1}
\! \! \! \left ( S_{\bt\bt} - \ioh \dth H_{\bt\bt} \right )
\psi^{\bt}_{\phs m} (t) \; ,
\eals
where we use the fact that $S^{\ct\bt}  S_{\bt\ct}=\one$
and $(A_{\ct\ct})^{-1} = (A^{-1})^{\ct\ct}$, while
$(A^{\ct}_{\phs\ct})^{-1} = (A^{-1})^{\ct}_{\phs\ct}$.
  In the first equation we find the conventional expression in
terms of the $\Srm^{-1}\Hrm$ matrix product, while the last one
shows a variant that does not require overlap inversion.
Although approximate in the evolution, it can be shown to be 
strictly unitary for finite $\dt$ (see Appendix~\ref{app:static-cn}).

\subsection{Constant connection}

  For a situation in which both the basis and the (tangent)
Hilbert space $\Omega(t)$ do change with time, we 
consider now the case in which $\dt$ is finite but small
enough so that the connection $D^{\mu}_{\phe\nu t}$  
can be taken as constant (the situation for varying basis set 
but within an invariant or converged $\Omega$ was contemplated 
in Ref.~\cite{Artacho2017}).

\subsubsection{Metric tensor evolution under constant connection}

  First,  let us see how the metric tensors evolve between 
$t$ and $t+\dt$. 
  \tcolr{In general, and still exact,}
\beq
\label{eq:diffeq-overlap}
\partial_t S_{\mu\nu} = D_{\mu t \nu} + D_{\mu\nu t} 
= D_{\mu t}^{\phm\sigma} S_{\sigma\nu} + 
S_{\mu\lambda} D^{\lambda}_{\phe\nu t} \, .
\eeq 
\tcolr{If} both $D^{\mu}_{\phe\nu t}$ and $D_{\mu t}^{\phm\nu}$
are \tcolr{taken as} constant, given the premise of this Section, and 
\tcolr{given} the fact that $D_{\mu t}^{\phm\nu} = D^{\nu\phm *}_{\phe\mu t}$,
the solution of Eq.~\eqref{eq:diffeq-overlap} is
\beq
\label{eq:evol-overlap}
S_{\mu\nu}(t+\dt) = e^{\dt D_{\mu t}^{\phm\sigma}}
S_{\sigma\lambda} (t) e^{\dt D^{\lambda}_{\phe\nu t}} \; ,
\eeq
\tcolr{which will be exact as long as those connections
are strictly constant, but will represent an approximate
solution for small $\dt$ but varying connections.}
  Analogously,
\beqs
\partial_t S^{\mu\nu} = D^{\mu\phs\nu}_{\phe t} + D^{\mu\nu}_{\phm t} 
= D^{\mu}_{\phe t \sigma} S^{\sigma\nu} + 
S^{\mu\lambda} D_{\lambda\phe t}^{\phe\nu} \, ,
\eeqs 
gives
\beq
\label{eq:evol-overlap-inv}
S^{\mu\nu}(t+\dt) = e^{\dt D^{\mu}_{\phe t \sigma}} \,
S^{\sigma\lambda} (t) \, e^{\dt D_{\lambda\phe t}^{\phe\nu}} \; ,
\eeq
since both $D^{\mu}_{\phe t \sigma}$ and $D_{\lambda\phe t}^{\phe\nu}$
are constant again.

\subsubsection{Calculation of the connection}
\label{sec:define-connect}

  The results in Eq.~\eqref{eq:evol-overlap} and
Eq.~\eqref{eq:evol-overlap-inv} are important, not only
for further algebraic manipulations, but because they represent
consistency conditions for the evolution, and, to some extent,
they define the connection.
  In explicit calculations, the overlap matrix is defined extrinsically
at any time step, i.e., it does not arise from evolution, but
given the positions of atoms at a given time step and given
the basis set definition in the larger ambient Hilbert space, as
\beq
\label{eq:explicit-overlap}
S_{\mu\nu}(t) = \int \! \phi^*_{\mu} (\mathbf{r},t) \phi_{\nu}(\mathbf{r},t) 
\, \mathrm{d}^3 \mathbf{r}
\eeq
where, typically, 
$\phi_{\mu}(\mathbf{r}, t)=\langle \mathbf{r} | e_{\mu}, t\rangle = 
\phi_{\mu} [\mathbf{r}- \mathbf{R}_{\mu}(t)]$, and where $\mathbf{R}_{\mu}(t)$ 
represents the position of the centre of the orbital at that time. 

  The connection itself can be computed extrinsically, normally as
\beqs
D^{\mu}_{\phs\nu t} = S^{\mu\sigma} D_{\sigma \nu t}
\eeqs
and 
\beq
\label{eq:explicit-connection}
D_{\sigma\nu t} = \int \! \phi^*_{\sigma} (\mathbf{r},t) \,
\partial_t \phi_{\nu}(\mathbf{r},t) \, \mathrm{d}^3 \mathbf{r} \; .
\eeq

  There are therefore two possibilities for an approximate evolution for 
finite $\dt$. ($i$) The connection can be calculated as in Eq.~\eqref{eq:explicit-connection}
neglecting the small discrepancy in the evolved overlap $S_{\mu\nu}(t+\dt)$
between the actual one, as in Eq.~\eqref{eq:explicit-overlap}, and the one
that would result from evolving under the calculated connection, as in
Eq.~\eqref{eq:evol-overlap}.
  Alternatively, ($ii$) a connection can be proposed, instead of via 
Eq.~\eqref{eq:explicit-connection}, by construction to satisfy Eq.~\eqref{eq:evol-overlap}
from the overlaps calculated using Eq.~\eqref{eq:explicit-overlap} both
at $t$ and $t+\dt$.

  In this work we test option $(i)$, since it is the most attractive numerically.
  The calculation of the connection via Eq.~\eqref{eq:explicit-connection} can be
done in linear-scaling operations.
  We have implemented it in the {\sc Siesta} program 
\cite{Soler2002, Artacho2008, Garcia2020}, 
which uses \tcolr{finite-support} atomic orbitals as basis sets, and 
contains a real-time TD-DFT implementation \cite{Tsolakidis2002}.
  The integrals in Eq.~\eqref{eq:explicit-connection} represent two-centre integrals
which are evaluated by a trivial extension of the method explained in
Section 5 of Ref.~\cite{Soler2002}. 
  The implemented connection is tested below. 
  Option ($ii$) is further explored in Appendix~\ref{app:option2}, where a possible
alternative direction towards better orthonormality preservation is outlined, 
albeit with worse scaling in computational expense as far as we can see.

\subsection{Unitarity and convergence with time step}

\subsubsection{Exponential evolution}
\label{sec:exponential_evolution}

  For constant  $D^{\mu}_{\phantom{e}\nu t}$ and $H^{\mu}_{\phantom{e}\nu}$
the coefficients for the ket and the bra evolve 
as given by the solution of Eqs.~ \eqref{eq:evoldiff} and \eqref{eq:evoldiffbra}, 
namely, 
\bal
\label{eq:ketevol}
\psi^{\mu}_{\phs m} (t+\dt) &= e^{- \dt 
(\ioh \, H^{\mu}_{\phe\nu} + D^{\mu}_{\phe\nu t})} \psi^{\nu}_{\phs m} (t) \\ 
\label{eq:ketevol2}
\psi_{m\mu} (t+\dt) &= \psi_{m\nu} (t) e^{ \dt (\ioh \, H^{\mu}_{\phe\nu} 
+ D^{\mu}_{\phe\nu t})} \; .
\eal
Checking again for unitarity,
\bals
\psi_{m\mu} & (t+\dt) \, \psi^{\mu}_{\phs n} (t+\dt) = \\
& \psi_{m\nu} (t) \, e^{ \dt \, (\ioh H^{\nu}_{\phe\mu} 
    + D^{\nu}_{\phe \mu t})}
e^{- \dt (\ioh \, H^{\mu}_{\phe\delta} 
+ D^{\mu}_{\phe\delta t})} \psi^{\delta}_{\phs n} (t) \; .
\eals
  Since $e^A e^B = e^{A+B}$ when $[A,B]=0$, then
\beqs
e^{ \dt \, (\ioh H^{\nu}_{\phe\mu} + D^{\nu}_{\phe\mu t})}
e^{- \dt (\ioh  H^{\mu}_{\phe\delta} + D^{\mu}_{\phe\delta t}) }= 
\delta^{\nu}_{\phs\delta} \,
\eeqs
and unitarity of the propagation in Eqs.~\eqref{eq:ketevol} and
\eqref{eq:ketevol2} is confirmed.
  Appendix~\ref{app:braevol-constconnect} shows that the bra evolved as in 
Eq.~\eqref{eq:ketevol2} is the instantaneous bra of the ket in
Eq.~\eqref{eq:ketevol}.
  In Appendix~\ref{app:parallel} the situation for parallel transport is
presented for clarity.

\subsubsection{Crank Nicolson}
\label{sec:CN}

  Unlike the static-basis case, the Crank Nicolson approximate 
evolution of Eqs.~\eqref{eq:ketevol} and \eqref{eq:ketevol2} is {\it not} 
unitary regardless of $\dt$.
  In order to see this let us proceed as follows.
  The Crank Nicolson algorithm for both equations is expressed as
\bal
\label{eq:cn-nonparallel}
\psi^{\ct}_{\phs m} (t+\dt) &= \left [ \one^{\ct}_{\phs\bt} + 
\dth \left \{ \ioh H^{\ct}_{\phs\bt} + D^{\ct}_{\phs\bt t} \right \} \right]^{-1} \times \\
& \nonumber
\quad \times 
\left ( \one^{\bt}_{\phs\bt} - \dth \left \{ \ioh H^{\bt}_{\phs\bt}  + 
D^{\bt}_{\phs\bt t} \right \} \right ) \psi^{\bt}_{\phs m} (t) \\
\nonumber
\psi_{m\ct} (t+\dt) &= \psi_{m\bt} (t)\left ( \one^{\bt}_{\phs\bt}  + 
\dth \left \{ \ioh H^{\bt}_{\phs\bt}  + D^{\bt}_{\phs\bt t} \right \}  \right) \times \\
&\label{eq:cn-nonparallel2}
\quad \times \left [ \one^{\bt}_{\phs\ct} - \dth 
\left \{ \ioh H^{\bt}_{\phs\ct} + D^{\bt}_{\phs\ct t} \right \} \right ]^{-1}  \; .
\eal
It is easy to see (using appropriate commutation) that, as defined, 
\beqs
\psi_{m\bt} (t+\dt) \psi^{\bt}_{\phs n} (t+\dt) = \psi_{m\bt} (t) \psi^{\bt}_{\phs n} (t) \; ,
\eeqs
and, therefore, scalar products would seem to be preserved exactly.

  It is not so, however. The evolved bra $\psi_{m\ct}(t+\dt)$ is only approximately 
the bra of the evolved ket $\psi^{\ct}_{\phs m}(t+\dt)$, and therefore unitarity will 
only be approximate.
  This is a curvature effect; Appendix~\ref{app:parallel} shows it to happen 
for parallel transport as well.
  The actual bra of the ket in Eq.~\eqref{eq:cn-nonparallel} is obtained by turning
it around and applying metric tensors as needed, as
\bal
\nonumber\psi_{m\ct}(2) &= \psi_{m\bt}(1) S^{\bt\bt}(1) 
\left ( \one_{\bt}^{\phe\bt} + \dth 
\left \{ \ioh H_{\bt}^{\phe\bt} - D_{\bt t}^{\phe\bt} \right \} \right ) \times \\ 
\label{eq:bra-cn-nonparallel}
& \quad \left [ \one_{\bt}^{\phe\bt} - \dth \left \{ \ioh H_{\bt}^{\phe\bt} - 
D_{\bt t}^{\phe\bt} \right \} \right ]^{-1} \! \! \! \! 
S_{\bt\ct}(2)
\eal
where we use 1 for $t$ and 2 for $t+\dt$. 
  Using Eq.~\eqref{eq:evol-overlap} for $S_{\ct\ct}(2)$ does not convert
Eq.~\eqref{eq:bra-cn-nonparallel} into Eq.~\eqref{eq:cn-nonparallel2}.
  The unitarity of the evolution is only approximate. 
  It is tested below.


\subsubsection{Tests: collision of two He atoms and H across graphite}


\begin{figure}[t] 
\includegraphics[width=0.4\textwidth]{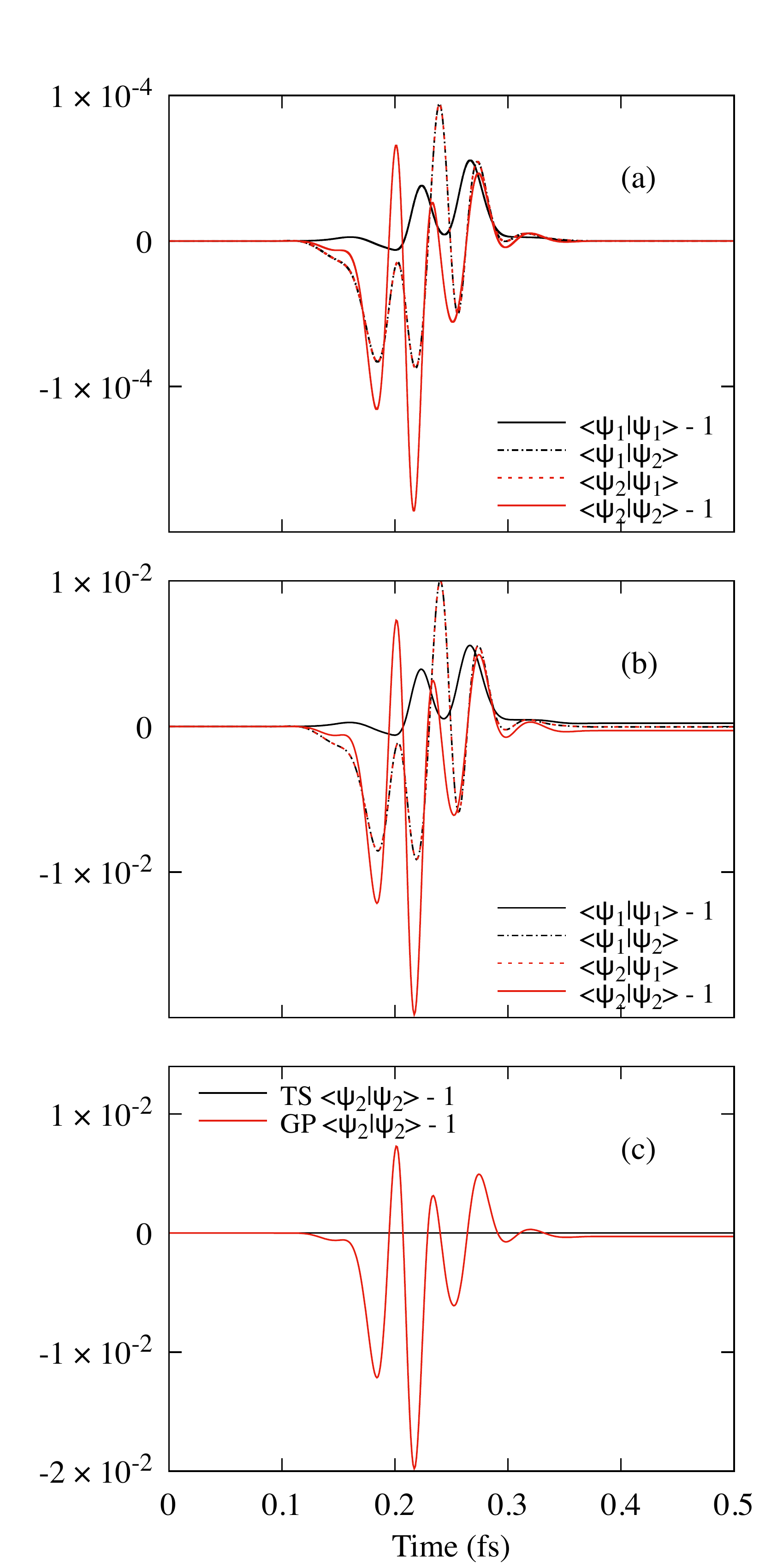}
\caption{Evolution of the deviation from orthonormality for 
the two occupied states for a collision between two He atoms
\tcolr{using the gauge-potential (GP) integrator},
comparing a timestep of (a) 0.01 as and (b) 1 as. 
(c) compares the evolution of $\langle \psi_2 | \psi_2\rangle -1$ 
for the two integrators, \tcolr{GP and Tomfohr-Sankey (TS)},
with $\dt$ = 1 as.}
\label{fig:orthonormality}
\end{figure} 


  The mentioned non-unitary evolution is explicitly shown numerically in
Fig.~\ref{fig:orthonormality}. 
  For the two occupied states, $|\psi_1\rangle$ and $|\psi_2\rangle$ in a collision 
between two He atoms, the quantities $\langle \psi_1 | \psi_1\rangle -1$, 
$\langle \psi_2 | \psi_2\rangle -1$, $\langle \psi_1 | \psi_2\rangle$, and
$\langle \psi_2 | \psi_1\rangle$ are plotted versus time for various values
of $\dt$, using the Crank Nicolson algorithm proposed in Eq.~\eqref{eq:cn-nonparallel}.
  One He atom is kept fixed in space while another moves past it on a fixed
trajectory with an impact parameter of 0.5 \AA, with a fixed velocity of
1 atomic unit.
  
  The calculations are performed using the {\sc Siesta} program, with 
a double-$\zeta$ polarised basis set.
  All the technical settings of the calculations are as in Ref.~\cite{Halliday2019},
for a box size of 10 \AA.
  It is apparent how the overlap matrix for the two evolving states deviates
from the starting unit matrix, depending on the size of $\dt$ as expected
from the discussion above.

  Using as reference \tcolr{the TS} algorithm \cite{Sankey}, which is unitary by 
construction and was extensively used in first-principles electronic stopping 
power calculations \cite{Zeb2012,Zeb2013, Ullah2015, Halliday2019, Gu2020}, 
the deterioration of orthonormality \tcolr{of the GP integrator} is
\tcolr{assessed by comparing both} for exactly the same process and 
approximations (using the same basis).
  Fig.~\ref{fig:orthonormality}(c) clearly shows the creeping in of deviation from 
orthonormality of the evolving states for the \tcolr{GP} algorithm of 
Eq.~\eqref{eq:cn-nonparallel} as compared with the strictly unitary \tcolr{TS}
alternative, which is limited only by the accuracy in the diagonalization involved. 
  The significant deviation between 0.1 and 0.35 fs is over the period 
where the two atoms are close enough to interact, i.e., when 
the basis states associated to the different atoms overlap.
  The magnitude of these larger deviations depends directly on the value 
of $\dt$, although the shape is identical, as can be seen by comparing 
Figs. \ref{fig:orthonormality}(a) and (b), but they return rapidly to close to 
zero once the atoms are further apart, with the final deviation from 0 at 
0.5 fs also depending on $\dt$ at a much smaller scale.


\begin{figure}[t] 
\includegraphics[width=0.4\textwidth]{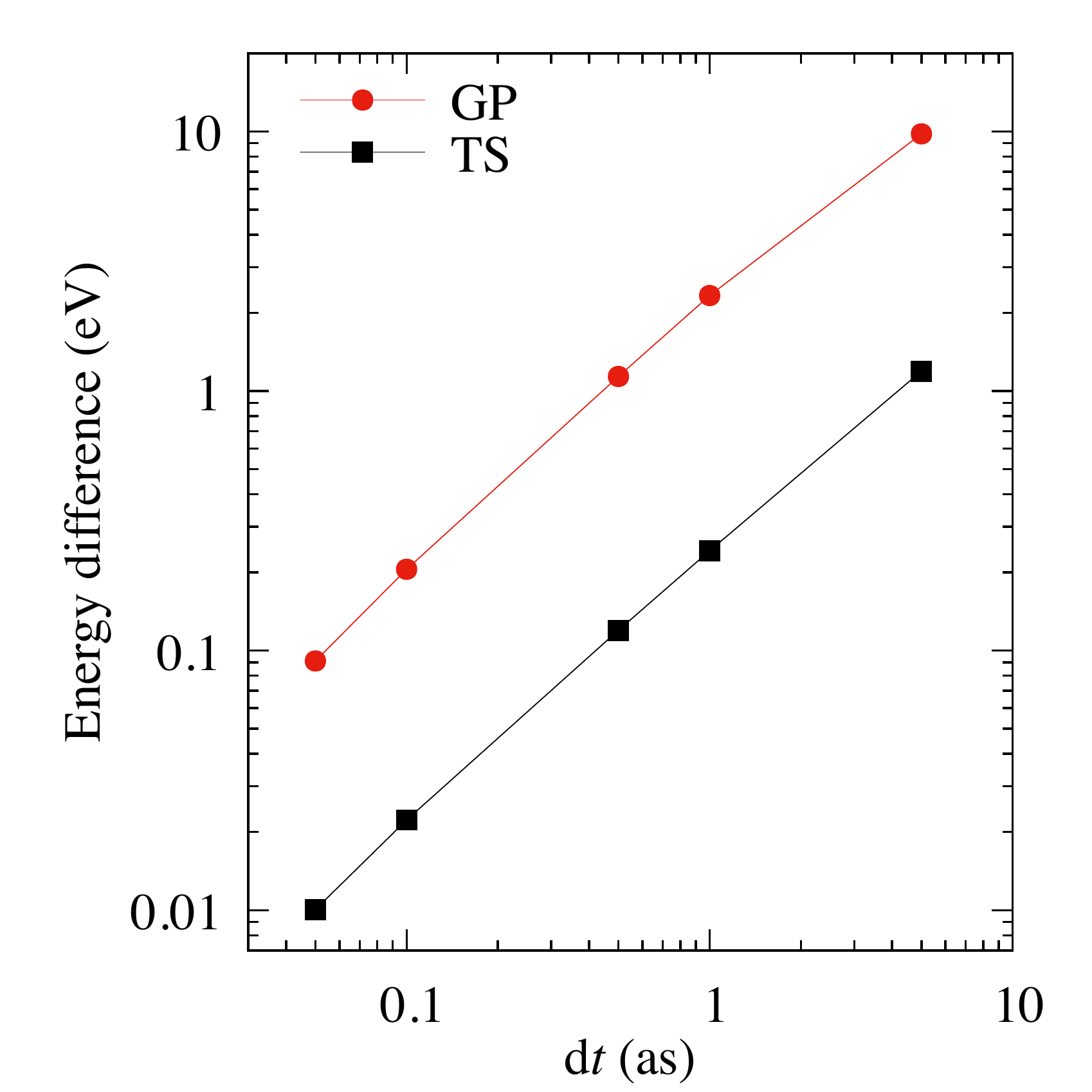}
\caption{Deviation in electronic energy uptake versus time step $\dt$ for
the collision between two He atoms, relative to $\dt$ = 0.01 as,
\tcolr{for both the TS and GP integrators}.}
\label{fig:energy-dt}
\end{figure} 


  That deviation from unitary evolution is behind the demand for smaller 
$\dt$ of the algorithm apparent in Fig.~\ref{fig:energy-dt}, where the 
convergence in energy transfer for the collision is shown
(difference in electronic energy between times before the collision
and after the collision).

\begin{figure}[t] 
\includegraphics[width=0.4\textwidth]{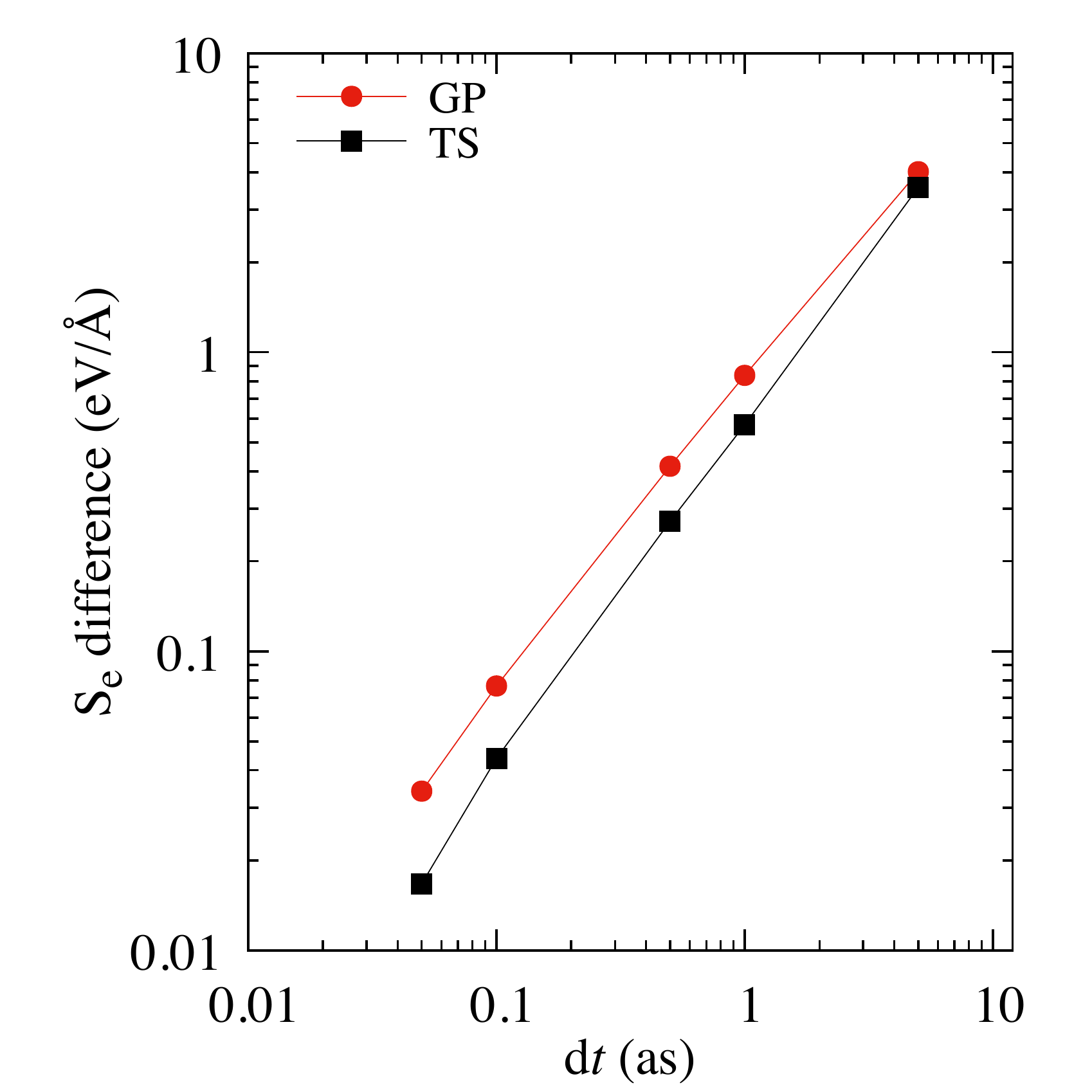}
\caption{Deviation in electronic stopping power $S_e$ versus time 
step $\dt$ for a proton travelling through graphite along the (0001) 
direction, \tcolr{for both the TS and GP integrators}.}
\label{fig:stopping-dt}
\end{figure}

  The convergence, however, depends on system and property.
  Fig.~\ref{fig:stopping-dt} shows an analogous plot for the electronic
stopping power $S_e$, electronic energy uptake per unit length 
traversed by a proton projectile travelling across graphite along
the (0001) direction (the details about this simulation can be found 
in Ref.~\cite{Halliday2019}).
  The figure shows that this property is quite similarly converged  
with both integrators.

  It should be noted, however, that the difference in computational effort
for both integrators is substantial.
  The \tcolr{GP} integrator following Eq.~\eqref{eq:cn-nonparallel}, requires the
calculation of the connection $D_{\mu\nu t}$, which represents two-centre
integrals that are pre-calculated as tables at the beginning of 
a simulation, which then are interpolated and rotated as needed.
  It represents a very small part of the {\sc Siesta} run, 
as the other two-centre integrals, such as the overlap and 
kinetic energy matrices.
  The \tcolr{TS} integrator requires a diagonalization of the 
overlap matrix for the whole basis at every time step.
  The difference increases with system size, since the effort to
calculation of $D_{\mu\nu t}$ scales linearly with system size 
(due to the sparsity of the matrix - same as for the overlap)
while the said diagonalization scales with the cube power.


\begin{figure}[t] 
\includegraphics[width=0.4\textwidth]{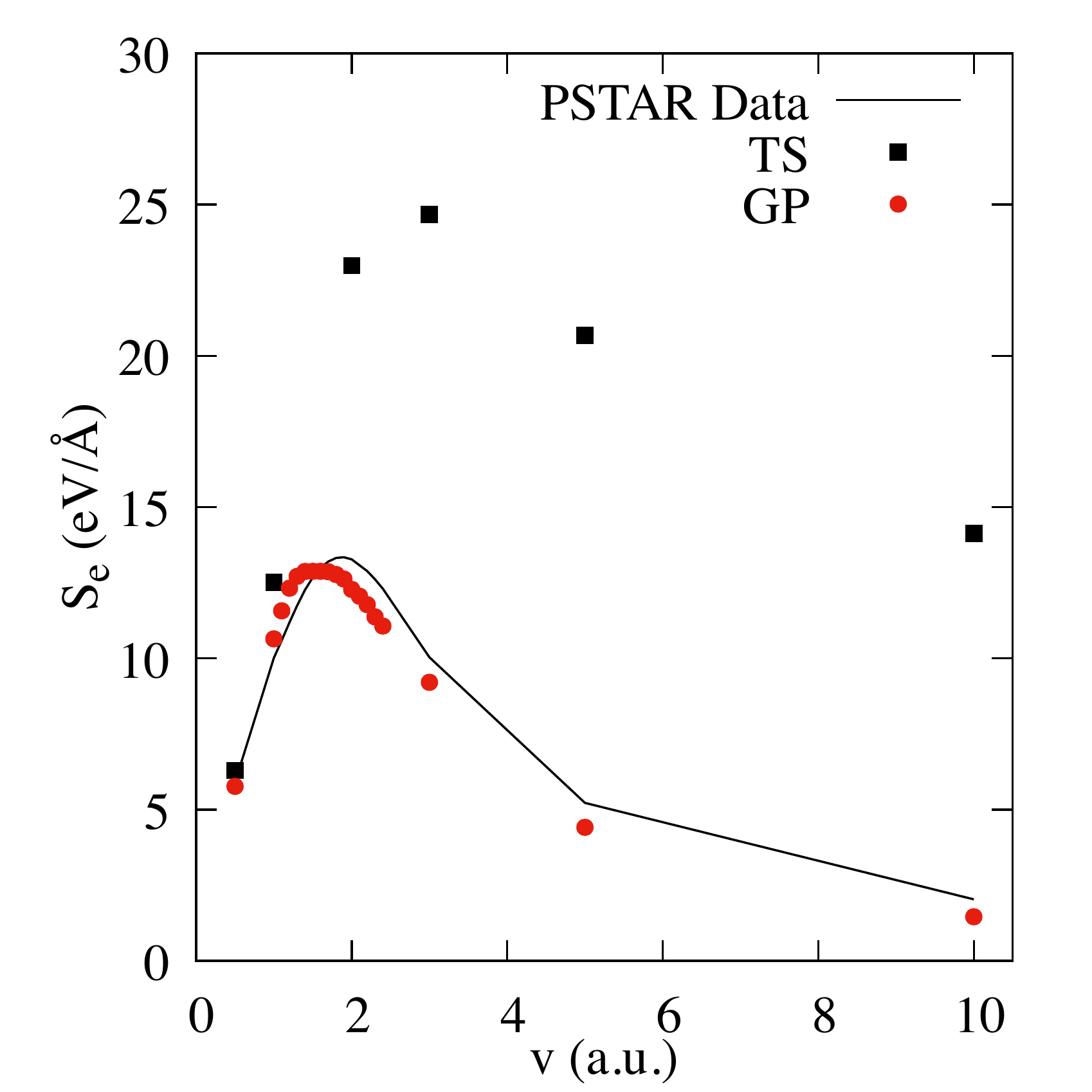}
\caption{Comparison of converged results 
\tcolr{for both the GP and TS integrators}, for the 
electronic stopping power $S_e$ versus velocity 
for a proton traveling through graphite along the (0001) 
direction. 
  PSTAR data are shown for comparison
\cite{PSTAR}.}
\label{fig:stopping}
\end{figure}

  It is important to finish, however, revisiting the accuracy of the 
converged integration with both algorithms.
  It was already pointed out in Ref.~\cite{Artacho2017} that the
\tcolr{TS} integrator, although perfectly unitary and 
displaying good convergence, does not produce the correct
integration for high velocities.
  It worked well for electronic stopping power calculations  for
low velocity projectiles \cite{Zeb2012, Zeb2013, Ullah2015, 
Halliday2019, Gu2020}.
  This is confirmed in Fig.~\ref{fig:stopping}, where the electronic
stopping power for a proton across graphite is depicted, 
comparing both integrators with the empirical PSTAR data
\cite{PSTAR}.
  $S_e$ is indeed well reproduced by both algorithms for
velocities below 1 atomic unit.


  However, the deviation at higher velocities for the \tcolr{TS}
integrator is very apparent, very clearly confirming the formal results of 
Ref.~\cite{Artacho2017}, with a large overestimation of the electronic
stopping power, by a factor of two around the Bragg peak, growing to
a tenfold overestimation for velocities around ten atomic units.

\begin{figure}[t] 
\includegraphics[width=0.4\textwidth]{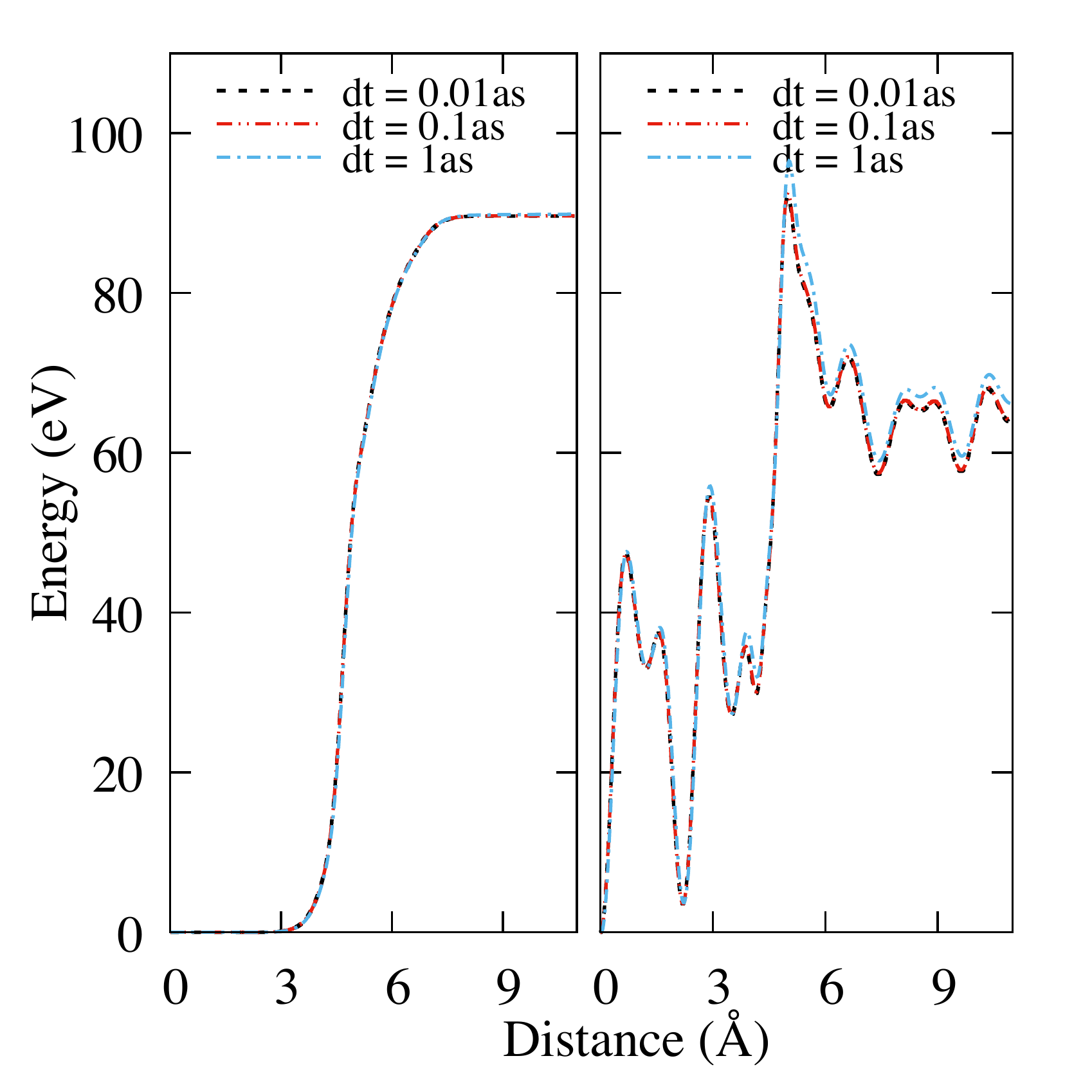}
\caption{Electronic energy as a function of position along the projectile
trajectory for a collision between two He atoms, one travelling at 1 a.u.
\tcolr{The left (right) panel shows the results for the TS (GP)
integrator.}}
\label{fig:2He-energy}
\end{figure}

  The simpler system of the two He atom collision is quite illustrative.
Fig.~\ref{fig:2He-energy} shows the electronic energy as a function of
position of the projectile He atom, as it passes by an immobile one.
  The \tcolr{TS} integrator converges better with $\dt$, as
shown in Fig.~\ref{fig:energy-dt}, but quite a lot of the physics is lost.
  In this case the key difference stems from the start of the evolution,
which is an abrupt kick of the projectile nucleus.
  It takes a time for the electrons around that nucleus to respond,
and there is a lag. 
  Since there is nothing else until reaching the target He atom, the 
electronic cloud around the projectile oscillates back and forth, 
which becomes a more complex behaviour when colliding.
  This is completely missed by the \tcolr{TS} algorithm,
since it transposes the coefficients of the evolving states 
from the (orthogonalised) basis at $t$ to the one at $t+\dt$,
which implies an instantaneous response to the initial kick,
without oscillation. 
  Very smooth, nicely converged, and quite unphysical.
  Admittedly, it is an example particularly ill-suited for the
algorithm, but illustrative, nevertheless.


\subsubsection{Orthonormalisation correction}

  Of course, the Crank-Nicolson propagation step of \tcolr{GP}, in
Eq.~\eqref{eq:cn-nonparallel}, can be made strictly unitary by force
if adding a L\"owdin orthonormalisation step using
$\mathcal{S}^{-1/2}(t+\dt)$ as obtained from the 
diagonalization of 
\beqs
\mathcal{S}_{mn} (t+\dt)= \langle \psi_m(t+\dt)|\psi_n(t+\dt)\rangle \, .
\eeqs
  Notice that the $\mathcal{S}$ matrix is of $N\times N$ dimensions,
\tcolr{$N$ being} the number of occupied propagating states, much 
smaller than the number of basis states. 
  
  In practical terms, and given the satisfactory unitarity
achieved directly by Eq.~\eqref{eq:cn-nonparallel} for small $\dt$,
one can choose to evolve using the uncorrected algorithm for
some time, then evaluating the $\mathcal{S}$ matrix once 
every number steps $n_s$, and, whenever 
\tcolr{$\max_{mn} \{ |\mathcal{S}_{mn}-\delta_{mn}| \}$} exceeds 
some tolerance $\epsilon$, an orthonormalisation 
step would be performed as described.
  The algorithm will be then optimised by choosing the best
combination of $\dt$, $n_s$ and $\epsilon$, which will 
depend on the system under study.

  We have shown above how different problems have different 
demands for convergence, depending, for instance, on 
their evolution being dominated by the basis motion and 
the connection, or by the Hamiltonian itself, either the
evolution of the external potential, or the Hamiltonian
effective spectral width for the evolving states. 
  The interplay of those parameters can therefore vary 
quite substantially.
  The method would be nicely completed with a learning
algorithm to adjust those parameters dynamically.


\section{Conclusions}

  The effect of the time evolution of the basis set and of the Hilbert space
it spans is explored for the description of the time evolution of 
quantum states.
  The exploration is both formal and numerical, assessing the deterioration
of the conservation of scalar products of evolving states (unitarity of
the evolution) and its implication for convergence with time step in the 
integration of the dynamical equation by time discretisation.

  Formally, a Crank-Nicolson \tcolr{algorithm using the connection of 
the manifold as gauge potential (GP)} is shown to keep unitarity only 
approximately, unlike the integrator proposed in Ref.~\cite{Artacho2017} 
for a moving basis within a static Hilbert space, and unlike the \tcolr{TS} 
algorithm \cite{Sankey}, which is perfectly unitary regardless of $\dt$ 
size by construction and works well at low velocities 
\cite{Zeb2012, Zeb2013, Ullah2015, Halliday2019, Gu2020}, but was 
suggested to describe unphysical evolution at higher velocities.

  The most numerically convenient \tcolr{GP} integrator 
[that of Eq.~\eqref{eq:cn-nonparallel}]
is tried on two systems, a collision between two He atoms and 
the passing of a constant-velocity H projectile across graphite.
  The unitarity and $\dt$ convergence for that algorithm is compared 
with the \tcolr{TS} one, the latter displaying the expected better
convergence with $\dt$, requiring from two to ten times less time steps
for a given simulation time.
  That advantage is however offset by the more efficient (and better 
basis-size scaling) \tcolr{GP} algorithm of Eq.~\eqref{eq:cn-nonparallel},
which only demands, per time step, the calculation of a sparse matrix of 
two-centre integrals (in linear-scaling operations) instead of the overlap matrix
diagonalization of \tcolr{TS}.

  The deviation from the physical evolution of the \tcolr{TS}
integration is confirmed for nuclear (and basis function) velocities
comparable to or larger than the valence electron velocities. 
  A very significant overestimation of the Bragg peak (a factor of two 
in both height and position) is observed for the electronic stopping power of 
the proton shooting through graphite along the centre of a (0001) 
channel.
  And a very clear modification of the expected physics is observed for
the same integrator when describing the two-atom collision.

  Although other routes for unitary integrators are proposed in this
work, only the approximately unitary \tcolr{GP} algorithm of 
Eq.~\eqref{eq:cn-nonparallel} is found to be satisfactory.
  The work also provides a better perspective in the understanding
of unitary evolution of quantum states with evolving basis sets, in the 
context of curved manifolds.

\begin{acknowledgments}
  E. Artacho is grateful for discussions with Prof. Christos Tsagas 
on the possibility of using extrinsic curvature for integrating, and 
acknowledges funding from the Leverhulme Trust, under Research 
Project Grant No. RPG-2018-254, from the EU through the 
ElectronStopping Grant Number 333813, within the Marie-Curie 
CIG program, and by  the Research Executive Agency under the 
European Union's Horizon 2020 Research and Innovation programme 
(project ESC2RAD, grant agreement no. 776410).
  Funding from Spanish MINECO is also acknowledged, through 
grant FIS2015-64886-C5-1-P, and from Spanish MICINN
through grant PID2019-107338RB-C61 / AEI /10.13039 / 501100011033.
  J. Halliday would like to acknowledge the EPSRC Centre for Doctoral 
Training in Computational Methods for Materials Science for funding 
under grant number EP/L015552/1. 
  This work has been performed using resources provided by the
Cambridge Tier-2 system operated by the University of Cambridge 
Research Computing Service  funded by Engineering 
and Physical Sciences Research Council Tier-2 
(capital Grant No. EP/P020259/1), and DiRAC funding from the 
Science and Technology Facilities Council.
  We also acknowledge the Partnership for Advanced Computing 
in Europe, PRACE, for awarding us access to computational 
resources in Joliot-Curie at GENCI@CEA, France, under
EU-H2020 Grant No. 2019215186.
\end{acknowledgments}


\appendix


\section{Unitary evolution in matrix representation}
\label{app:unitary-matrix}

  It can be checked more conventionally, using for the bra coefficients 
the complex conjugate of the ones for the ket, 
$\psi_m^{\phn\mu}=\langle \psi_m | e^{\mu}\rangle=\psi^{\mu *}_{\phe m}$ , 
as follows:
\bals
\partial_t & \langle \psi_m | \psi_n \rangle 
= \partial_t (\psi_m^{\phn\nu} \, S_{\nu\mu} \,\psi^{\mu}_{\phe n}) \\
&=(\partial_t \psi_m^{\phn\nu}) S_{\nu\mu} \psi^{\mu}_{\phe n} +
\psi_m^{\phn\nu} (\partial_t S_{\nu\mu}) \psi^{\mu}_{\phe n} +
\psi_m^{\phn\nu} S_{\nu\mu} (\partial_t \psi^{\mu}_{\phe n}) \\
&= \psi_m^{\phn\bt} (\ioh H_{\bt}^{\phs\bt} - D_{\bt t}^{\phm\bt}) 
 S_{\bt\bt} \psi^{\bt}_{\phs n} 
+  \psi_m^{\phn\bt} (D_{\bt t \bt} + D_{\bt\bt t}) \psi^{\bt}_{\phs n}  -\\
& \qquad \qquad - \psi_m^{\phn\bt} S_{\bt\bt} 
(\ioh H^{\bt}_{\phs\bt} + D^{\bt}_{\phs\bt t}) \psi^{\bt}_{\phs n} \\
&= \psi_m^{\phn\bt} (\ioh H_{\bt\bt} - D_{\bt t\bt}) \psi^{\bt}_{\phs n} 
+  \psi_m^{\phn\bt} (D_{\bt t \bt} + D_{\bt\bt t}) \psi^{\bt}_{\phs n}  -\\
& \qquad \qquad - \psi_m^{\phn\bt}
(\ioh H_{\bt\bt} + D_{\bt\bt t}) \psi^{\bt}_{\phs n} = 0\; ,
\eals 
where a filled bullet has been introduced for every upper (lower)
index that contracts with a contiguous lower (upper) index, 
and where we have used the fact that
$\partial_t S_{\nu\mu} = D_{\mu t \nu} + D_{\mu\nu t}$.

\section{Non-orthogonal static basis}
\label{app:unitary-static}

\subsection{Exponential solution}

\subsubsection{In the matrix representation}
\label{app:static-exp}

  A more traditional proof than the one in Eq.~\eqref{eq:demo-static} 
is presented here for the unitarity of the evolution of states in a static 
Hilbert space $\Omega$, for a non-orthogonal static basis.
  It is well known, but it serves to set up the scene for later manipulations.
  The $ \psi^{\mu}_{\phe m}$ elements
are actually the conventional expansion coefficients of $|\psi_m\rangle$
in the $\{ |e_{\mu}\rangle\}$ basis, as $|\psi_m\rangle =  |e_{\mu}\rangle
\langle e^{\mu}| \psi_m\rangle = |e_{\mu}\rangle \psi^{\mu}_{\phe m}$.
  Scalar products remain constant:
\bals
\langle \psi_m(t+&\dt) | \psi_n (t+\dt) \rangle = 
\psi^{\mu\phe *}_{\phe m} (t+\dt) \, S_{\mu\nu} \psi^{\nu}_{\phe n} (t+\dt) \\
&= \psi^{\bt\phs *}_{\phs m} (t+\dt) \, S_{\bt\bt} \, \psi^{\bt}_{\phs n} (t+\dt) \\ 
&= \psi^{\bt\phs *}_{\phs m} (t) \, e^{\dt \ioh H_{\bt}^{\phs\bt}} S_{\bt\bt} 
\, e^{-\dt \ioh H^{\bt}_{\phs\bt}}\psi^{\bt}_{\phs n} (t) \\ 
&= \psi^{\bt\phs *}_{\phs m} (t) \, S_{\bt\bt} \, S^{\bt\bt} \,
e^{\dt \ioh H_{\bt}^{\phs\bt}} S_{\bt\bt} 
\, e^{-\dt \ioh H^{\bt}_{\phs\bt}}\psi^{\bt}_{\phs n} (t) \\ 
&= \psi^{\bt\phs *}_{\phs m} (t) \, S_{\bt\bt} \, \,
e^{\dt \ioh H^{\bt}_{\phs\bt}}  
\, e^{-\dt \ioh H^{\bt}_{\phs\bt}}\psi^{\bt}_{\phs n} (t) \\ 
&=  \psi^{\bt\phs *}_{\phs m} (t) \, S_{\bt\bt} \, \psi^{\bt}_{\phs n} (t)
= \langle \psi_m(t) | \psi_n (t) \rangle \; ,
\eals
%
%
where filled bullets indicate contracted indices that are contracted
as in Appendix~\ref{app:unitary-matrix}.
In the fourth line, 
$\one = S_{\bt\bt} \, S^{\bt\bt}$
was introduced, and in the fifth line we made use of the following
relationship
\beq
\label{eq:raise-exp}
S^{\bt\bt}  e^{A_{\bt}^{\phs\bt}} S_{\bt\bt} = 
e^{S^{\bt\bt} A_{\bt}^{\phs\bt} S_{\bt\bt}} =
e^{A^{\bt}_{\phs\bt}}
\eeq
  The first equality is easily checked by expanding the exponential, 
and using the fact that 
\beqs
S^{\bt\bt}  (A_{\bt}^{\phs\bt})^n S_{\bt\bt} =
(S^{\bt\bt}  A_{\bt}^{\phs\bt} S_{\bt\bt})^n \; ,
\eeqs
since $A_{\bt}^{\phs\bt} A_{\bt}^{\phs\bt} =
A_{\bt}^{\phs\bt} S_{\bt\bt}
S^{\bt\bt} A_{\bt}^{\phs\bt}$ in any 
product in the power expansion.
  Eq.~\eqref{eq:raise-exp} indicates how to change representation
in powers and exponentials of rank-two tensors, but only between
$A_{\bt}^{\phs\bt}$ and $A^{\bt}_{\phs\bt}$
(the inverse of Eq.~\eqref{eq:raise-exp}, 
$S_{\bt\bt}  e^{A^{\bt}_{\phs\bt}}
S^{\bt\bt} = e^{A_{\bt}^{\phs\bt}}$,
is also true).
  Powers or exponentials of $A_{\bt\bt}$ or 
$A^{\bt\bt}$ do not make sense.

  Although we are using tensors here, we have used the matrix representation
as the traditional one, where the vector of coefficients for the bra is the
Hermitian conjugate of those for the ket. 
  Traditionally, $H_{\bt}^{\phs\bt}$ appears as $H_{\bt\bt} \, S^{\bt\bt}$,
normally expressed using conventional matrices, $\Hrm \, \Srm^{-1}$, and
$H^{\bt}_{\phs\bt}= S^{\bt\bt} \, H_{\bt\bt} = \Srm^{-1} \, \Hrm$.
  In Section~\ref{sec:constHD} an easier proof is presented for the
natural representation, using $\psi_{m\bt}$ for the bra, and $H^{\bt}_{\phs\bt}$
for the Hamiltonian.

\subsubsection{Correspondence of evolved bra and ket}
\label{app:static-bra}

  It is easy to check that the bra coefficients in this representation,
in the lower equation of Eq.~\eqref{eq:ketbraevol0}, represent the
actual bra of the upper equation. 
  If $\psi^{\mu}_{\phe m} (t+\dt) = \langle e^{\mu}|\psi_m(t+\dt)\rangle$,
turning it around gives
\beqs
\langle \psi_m(t+\dt) | e^{\mu}\rangle = \psi^{\mu\phe *}_{\phe m} (t+\dt)
= \psi^{\nu\phe *}_{\phe m} (t) e^{\dt \frac{i}{h} H_{\nu}^{\phe\mu}} \; ,
\eeqs
where we have just complex-conjugated the whole equation, 
and changed the order of $\psi$ and the exponential to 
reflect the index contraction ($\nu$) as the order in a matrix product.
  The same equation can be re-expressed as 
\beqs
 \psi_{m\sigma} (t+\dt) S^{\sigma\mu}
= \psi_{m\lambda} (t) S^{\lambda\nu} e^{\dt \frac{i}{h} H_{\nu}^{\phe\mu}} \; ,
\eeqs
or,  multiplying by the overlap on the right,
\beqs
 \psi_{m\mu} (t+\dt) 
= \psi_{m\lambda} (t) S^{\lambda\nu} e^{\dt \frac{i}{h} H_{\nu}^{\phe\sigma}} 
S_{\sigma\mu} \; ,
\eeqs
which becomes 
\beqs
 \psi_{m\mu} (t+\dt) 
= \psi_{m\nu} (t)  e^{\dt \ioh H^{\nu}_{\phe\mu}}  \; ,
\eeqs
by virtue of Eq.~\eqref{eq:raise-exp}, which coincides with
the second line in Eq.~\eqref{eq:ketbraevol0}, confirming the
expectation that the bra of the evolved ket coincides with
the evolved bra.

\subsection{Crank Nicolson unitarity for static basis}
\label{app:static-cn}

  In the natural representation, the bra for the ket in Eq.~\eqref{eq:ket-cn-static}  is 
%
\beq
\label{eq:bra-cn}
\psi_{m\ct}(t+\dt) = 
\psi_{m\bt} (t) 
\left ( \one^{\bt}_{\phs\bt} + \ioh \dth H^{\bt}_{\phs\bt} \right )
\left [ \one^{\bt}_{\phs\ct} - \ioh \dth H^{\bt}_{\phs\ct} \right ]^{-1}
\eeq
and, therefore $\langle \psi_m | \psi_n\rangle = \psi_{m\mu} \psi^{\mu}_{\phe n}$
relate at different times as 
\bals
\psi_{m\bt} &(t+\dt)  \psi^{\bt}_{\phs n} (t+\dt) = \\
= & \; \psi_{m\bt} (t) 
\left ( \one^{\bt}_{\phs\bt} + \ioh \dth H^{\bt}_{\phs\bt} \right ) 
\left [ \one^{\bt}_{\phs\bt} - \ioh \dth H^{\bt}_{\phs\bt} \right ]^{-1} \times \\
& \times 
\left [ \one^{\bt}_{\phs\bt} + \ioh \dth H^{\bt}_{\phs\bt} \right ]^{-1} 
\left ( \one^{\bt}_{\phs\bt} - \ioh \dth H^{\bt}_{\phs\bt} \right ) 
\psi^{\bt}_{\phs n} (t) \\
= & \; \psi_{m\bt} (t) 
\left ( \one^{\bt}_{\phs\bt} + \ioh \dth H^{\bt}_{\phs\bt} \right ) 
\left [ \one^{\bt}_{\phs\bt} + \ioh \dth H^{\bt}_{\phs\bt} \right ]^{-1} \times \\
& \times 
\left [ \one^{\bt}_{\phs\bt} - \ioh \dth H^{\bt}_{\phs\bt} \right ]^{-1} 
\left ( \one^{\bt}_{\phs\bt} - \ioh \dth H^{\bt}_{\phs\bt} \right ) 
\psi^{\bt}_{\phs n} (t) \\
= & \; \psi_{m\bt} (t)  \psi^{\bt}_{\phs n} (t) \; .
\eals
We have made use of the fact that the two inverse terms
in the middle commute.

In the traditional matrix representation
$\langle \psi_m | \psi_n\rangle = \psi^{\mu\phe *}_{\phs m} 
S_{\mu\nu} \psi^{\nu}_{\phe n}$,
but since 
\beqs
\psi_{m\mu}(t+\dt)  = \psi^{\mu\phe *}_{\phs m} (t+\dt) S_{\mu\nu}  
\eeqs
it obviously complies, too. 
  If wanting to check for the evolved one,
\bals
&\psi^{\bt\phe *}_{\pht m} (t+\dt) \, S_{\bt\ct} = \\
& = \psi^{\bt\phe *}_{\pht m} (t) 
\left ( \one_{\bt}^{\pht\bt} + \dth \ioh H_{\bt}^{\pht\bt} \right )
\left [ \one_{\bt}^{\pht\bt} - \dth\ioh H _{\bt}^{\pht\bt} \right ]^{-1}
\! \! \! S_{\bt\ct} \\
& = \psi^{\bt\phe *}_{\pht m} (t)
S_{\bt\bt} S^{\bt\bt} \!
\left ( \one_{\bt}^{\pht\bt} + \dth \ioh H_{\bt}^{\pht\bt} \right )
\Srm_{\bt\bt} \Srm^{\bt\bt} \!
\left [ \one_{\bt}^{\pht\bt} - \dth\ioh H _{\bt}^{\pht\bt} \right ]^{-1}
\! \! \! \! \! \! S_{\bt\ct} \\
& = \psi^{\bt\phe *}_{\pht m} (t) S_{\bt\bt} 
\left ( \one^{\bt}_{\pht\bt} + \dth \ioh H^{\bt}_{\pht\bt} \right )
\left [ \one^{\bt}_{\pht\ct} - \dth\ioh H ^{\bt}_{\pht\ct} \right ]^{-1}
\eals
which coincides with Eq.~\eqref{eq:bra-cn}, and the proof of
unitarity follows from there.

\section{Parallel transport}
\label{app:parallel}

\subsection{Evolution}

  The dynamics expressed in Eqs.~\eqref{eq:evoldiff} and \eqref{eq:evoldiffbra}
is in some sense counterintuitive, given the fact that neither
$D^{\mu}_{\phantom{e}\nu t}$ nor $D_{\mu\nu t}$ are
antihermitian, and it is hard to see how and why they would 
give unitary propagation.  
  The parallel transport case can be illustrative,
i.e.,   the evolution of states for zero covariant derivative.
  In an Euclidean space the states would not change in time, but
they do in a curved manifold.
  Because of the Schr\"odinger equation, zero covariant derivative 
implies $H^{\mu}_{\phantom{e}\nu}(t) = 0$.
  Parallel transport for ket and bra would then be described by
\bal
\label{eq:evolparallel}
\partial_t \psi^{\mu}_{\phantom{e}m} &=  
-  D^{\mu}_{\phantom{e}\nu t}  \, \psi^{\nu}_{\phantom{e}m} \nonumber \\
\partial_t \psi_{m\mu} &= \phantom{-} \psi_{m\nu} 
D^{\nu}_{\phantom{e}\mu t}    \; .
\eal
  
  For constant  $D^{\mu}_{\phantom{e}\nu t}$ 
the coefficients for the ket and the bra evolve as given by
the solution of Eq.~ \ref{eq:evolparallel}, namely, 
\beqa
\label{eq:ketbraevol}
\psi^{\mu}_{\phantom{e}m} (t+\mathrm{d}t) &= e^{- \mathrm{d}t 
D^{\mu}_{\phantom{e}\nu t}} \, \psi^{\nu}_{\phantom{e}m} (t) \nonumber \\ 
\psi_{m\mu} (t+\mathrm{d}t) &= 
\psi_{m\nu} (t) \, e^{ \mathrm{d}t 
D^{\nu}_{\phantom{e}\mu t}} \; .
\eeqa

  In the $\Omega(t)$ manifold, the scalar-product-preserving propagation
$\langle \psi_m | \psi_n\rangle (t+\mathrm{d}t) = \langle \psi_m | \psi_n\rangle (t)$
becomes
\begin{equation}
\label{eq:unitary}
\psi_{m\mu} (t+\mathrm{d}t) \psi^{\mu}_{\phantom{e}n} (t+\mathrm{d}t)
= \psi_{m\mu} (t) \psi^{\mu}_{\phantom{e}n} (t) \, .
\end{equation}
  Using Eq.~\ref{eq:ketbraevol}, 
\beqs
\psi_{m\mu} (t+\mathrm{d}t) \psi^{\mu}_{\phantom{e}n} (t+\mathrm{d}t) = 
\psi_{m\nu} (t) e^{ \mathrm{d}t \,  D^{\nu}_{\phantom{e} \mu t}}
e^{- \mathrm{d}t D^{\mu}_{\phantom{e}\delta t}} \psi^{\delta}_{\phantom{e}n} (t) \; .
\eeqs
  Since 
$
e^{ \mathrm{d}t \, D^{\nu}_{\phantom{e} \mu t}}
e^{- \mathrm{d}t D^{\mu}_{\phantom{e}\delta t}}= \delta^{\nu}_{\phantom{e}\delta} \,
$,
the unitarity of propagation in Eq.~\eqref{eq:unitary} is demonstrated. The key
is in Eq.~\eqref{eq:signchange}.

  
\subsection{Evolved bra}  
  
  We have shown that the scalar product between the evolved representation
of the ket $\psi^{\mu}_{\phe m}(t) \in \Omega(t)$ and the evolved representation 
of its bra, $\psi^n_{\phe\nu}(t)$, is preserved at later times. 
  We can also show the preservation of the scalar product
directly for the bra $\psi_{n\nu}(t+\dt)$, corresponding to 
$\psi^{\nu}_{\phe n}(t+\dt) \in \Omega(t+\dt)$.
  In other words, the evolved bra is the bra of the evolved ket, i.e., 
we want to check that for the ket represented by
\beq
\label{eq:ket}
\psi^{\mu}_{\phantom{e}m}(t+\dt) = e^{-\dt D^{\mu}_{\phantom{e}\nu t}}
\, \psi^{\nu}_{\phantom{e}m}(t) \; ,
\eeq
the bra would be
\beq
\label{eq:bra-evol}
\psi_{m\mu}(t+\dt) = \psi_{m\nu}(t)
\, e^{\dt D^{\nu}_{\phantom{e}\mu t}} \; .
\eeq
For the purpose let us write down Eq.~\ref{eq:ket} as
\beqs
\langle e^{\mu}, t+\dt | \psi_m, t+\dt \rangle = 
e^{-\dt \langle e^{\mu} | \partial_t e_{\nu} \rangle }
\langle e^{\nu}, t | \psi_m, t \rangle 
\eeqs
and turn it around (and complex conjugate it)
\beq
\label{eq:backwards}
\langle \psi_m, t+\dt |  e^{\mu}, t+\dt\rangle = 
\langle \psi_m, t | e^{\nu}, t \rangle 
\, e^{-\dt \langle \partial_t e_{\nu} |  e^{\mu} \rangle } \, , 
\eeq
where we have changed the order of factors in the 
r.h.s. to reflect the summed index $\nu$.
  Now, 
\beqs
\langle \psi_m |  e^{\mu} \rangle = 
\psi_{m\sigma} \, S^{\sigma\mu} \; ,
\eeqs
and Eq.~\ref{eq:backwards} becomes
\beqs
\psi_{m\sigma} (t+\dt) \, S^{\sigma\mu} (t+\dt) =
\psi_{m\lambda} (t) \, S^{\lambda\nu} (t)
e^{-\dt D_{\nu t}^{\phantom{m}\mu} } \, .
\eeqs
or 
\beqs
\psi_{m\mu} (t+\dt)  =
\psi_{m\lambda} (t) \, S^{\lambda\nu} (t)
e^{-\dt D_{\nu t}^{\phantom{m}\sigma} } \, S_{\sigma\mu} (t+\dt) \, .
\eeqs
Using now  
$S_{\ct\ct}(t+\dt)=e^{\dt D_{\ct t}^{\phn\bt}}S_{\bt\bt}(t)
e^{\dt D^{\bt}_{\phs\ct t}}$ [Eq.~\eqref{eq:evol-overlap}],
\bals
\psi_{m\mu} &(t+\dt)  = \\
&= \psi_{m\lambda} (t) \, S^{\lambda\nu} (t)
e^{-\dt D_{\nu t}^{\phantom{m}\sigma} } \, 
e^{\dt D_{\sigma t}^{\phm\delta}}S_{\delta\kappa} (t) 
e^{\dt D^{\kappa}_{\phe\mu t}} \\
&= \psi_{m\lambda} (t) \, S^{\lambda\nu} (t)
S_{\nu\kappa} (t) 
e^{\dt D^{\kappa}_{\phe\mu t}} \\
&= \psi_{m\lambda} (t) \, e^{\dt D^{\lambda}_{\phe\mu t}} \, .
\eals
coinciding with the evolved bra in Eq.~\ref{eq:bra-evol},
as expected.
  We have used that 
$e^{-\dt D_{\bt t}^{\phn\bt}} \, e^{\dt D_{\bt t}^{\phn\bt}} = 
e^{\dt ( D_{\bt t}^{\phn\bt} - D_{\bt t}^{\phn\bt})} = \one$
and that $S^{\ct\bt} \, S_{\bt\ct}=\one^{\ct}_{\phs\ct}$.

\subsection{Crank Nicolson for parallel transport}

  Above we just saw that since parallel transport preserves scalar 
products, it also does it in the case of a constant connection, no matter for
how long.
  The question is now what happens with approximate evolution 
over a finite $\dt$. 

  Since the evolution of both ket and bra under parallel transport with
constant connection [Eq.~\eqref{eq:ketbraevol}] is mathematically 
analogous to their evolution under Eq.~\eqref{eq:ketbraevol0} with 
fixed basis, the Crank Nicolson algorithm can be used to approximate
their transport,
\bal
\label{eq:cn-parallel}
\psi^{\ct}_{\phs m} (t+\dt) &= \left [ \one^{\ct}_{\phs\bt} + \dth D^{\ct}_{\phs\bt t} 
\right]^{-1} \! \! \!
\left ( \one^{\bt}_{\phs\bt}  - \dth D^{\bt}_{\phs\bt t} \right ) \psi^{\bt}_{\phs m} (t) \\
\label{eq:cn-parallel2}
\psi_{m\ct} (t+\dt) &= \psi_{m\bt} (t)\left ( \one^{\bt}_{\phs\bt} + \dth D^{\bt}_{\phs\bt t} 
\right) \! \! \!
\left [ \one^{\bt}_{\phs\ct} - \dth D^{\bt}_{\phs\ct t} \right ]^{-1}  \; .
\eal
  As done in Section~\ref{sec:CN},
it is apparent (using appropriate commutation) that
\beqs
\psi_{m\bt} (t+\dt) \psi^{\bt}_{\phs n} (t+\dt) = \psi_{m\bt} (t) \psi^{\bt}_{\phs n} (t) \; ,
\eeqs
and, therefore, scalar products would seem to be preserved exactly.
But again, they are not.
  The evolved bra $\psi_{m\ct}(t+\dt)$ as in 
Eq.~\eqref{eq:cn-parallel2} is only approximately the bra of the evolved ket
$\psi^{\ct}_{\phs m}(t+\dt)$, and therefore unitarity will only be approximate.
  The actual bra of the ket in Eq.~\eqref{eq:cn-parallel} is rather
\bal
\nonumber
\psi_{m\ct}(t+\dt) & = \psi_{m\bt}(t) S^{\bt\bt}(t) 
\left ( \one_{\bt}^{\phs\bt} - \dth D_{\bt t}^{\phe\bt} \right ) \times \\ \label{eq:bra-cn-parallel}
&\quad \times \left [ \one_{\bt}^{\phs\bt} + \dth D_{\bt t}^{\phe\bt} \right ]^{-1} \! \! \! \! 
S_{\bt\ct}(t+\dt) \; .
\eal
  Using Eq.~\eqref{eq:evol-overlap} for $S_{\ct\ct}(t+\dt)$ does not convert
Eq.~\eqref{eq:bra-cn-parallel} into Eq.~\eqref{eq:cn-parallel2}.



\section{Alternative connection and integrator for finite $\dt$}
\label{app:option2}

  When travelling between $t$ and $t+\dt$,  
$S_{\mu\nu}(t)$ and $S_{\mu\nu}(t+\dt)$ can be calculated 
directly in ambient space using Eq.~\eqref{eq:explicit-overlap}.
  Assuming a constant connection we could then use
the $D_{\mu\nu t}$ arising from the solution of Eq.~\ref{eq:evol-overlap}, 
instead of calculating it explicitly as in Eq.~\eqref{eq:explicit-connection}.
  However, the relation between overlaps alone should
not be sufficient for the determination of the connection,
since any rotation of the basis at $t+\dt$ should leave 
$S_{\mu\nu}(t+\dt)$ unaltered while the transformation
between $t$ and $t+\dt$ would change.

\subsection{Parallel transport transformation}

  Defining the transformation tensor,
\beq
\label{eq:def-transf}
\cA^{\mu}_{\phs\nu} = e^{-\dt D^{\mu}_{\phs\nu t}} \; .
\eeq
it performs the transformation from $t$ to $t+\dt$ for any 
vector following parallel transport (see Appendix~\ref{app:parallel}),
\beq
\label{eq:finite-parallel}
\psi^{\mu}_{\phs m}(t+\dt) = \cA^{\mu}_{\phs\nu} \psi^{\nu}_{\phs m}(t) \, .
\eeq
  We could define $\cA^{\ct}_{\phs\ct}$ more generally as
the one doing that operation regardless of $D^{\mu}_{\phs\nu t}$
being constant or not. 
  If not, Eq.~\eqref{eq:def-transf} would have to be replaced by a 
time-ordered integral, but once in possession of $\cA^{\mu}_{\phs\nu}$,
the rest of this section would be the same.

It is analogous to a basis set transformation, except that 
$\cA^{\mu}_{\phe\nu} \neq A^{\mu}_{\phe\nu} \equiv 
\langle e^{\mu}, t+\dt \, | \, e_{\nu},t\rangle$, 
as defined in Ref.~\cite{Artacho2017}, which is 
the relevant transformation when $\Omega$ does
not change with time.

  The parallel-transport transformation given by 
Eq.~\eqref{eq:finite-parallel} can be understood  as a basis
set transformation within $\Omega(t+\dt)$ between
the basis set $\{|e_{\mu},t\rangle\}$ of $\Omega(t)$
parallel-transported into $\Omega(t+\dt)$ and the
actual basis of the latter space, $\{|e_{\mu},t+\dt\rangle\}$.
  Define $|\phi_{\mu}, t+\dt\rangle$ as the parallel transport 
onto $\Omega(t+\dt)$ of $|e_{\mu},t\rangle$.
  Following Eq.~\eqref{eq:finite-parallel}, its expansion in the 
basis of $\Omega(t+\dt)$ would be
\beqs
\phi^{\nu}_{\phe\mu} (t+\dt) = \cA^{\nu}_{\phe\sigma} 
\phi^{\sigma}_{\phe\mu} (t)
\eeqs
but $\phi^{\sigma}_{\phe\mu} (t) = \delta^{\sigma}_{\phe\mu}$
(since $|\phi_{\mu}, t \rangle = |e_{\mu}, t\rangle$), and, therefore,
\beq
\label{eq:basistransf}
\phi^{\nu}_{\phe\mu} (t+\dt) = \cA^{\nu}_{\phe\mu} \; ,
\eeq 
that is, $\cA^{\nu}_{\phe\mu} = \langle e^{\nu},t+\dt | 
\phi_{\mu}, t+\dt\rangle$, and $|\phi_{\mu},t+\dt\rangle =
|e_{\nu},t+\dt\rangle \, A^{\nu}_{\phe\mu}$, the proposed basis set 
transformation.

  Writing it with matrices, if defining matrix $\Arm$ as the 
tensor $\cA^{\ct}_{\phs\ct}$, parallel transport from $t$ to
$\dt$ becomes
\beqs
\Psi_m(t+\dt) = \Arm \, \Psi_m (t) \, ,
\eeqs
and the overlap evolution, Eq.~\ref{eq:evol-overlap} can be recast as 
\bal
\label{eq:def-exp-connect}
&\Srm(t) = \mathrm{A}^{\dagger} \Srm(t+\dt) \mathrm{A} \\
&\Srm(t+\dt) = (\Arm^{\dagger})^{-1}  \Srm(t) \Arm^{-1} \; ,
\eal
Eq.~\eqref{eq:evol-overlap-inv} for the other metric tensor
$\Srm^{\ct\ct}$ becomes
\bal
&\Srm(t)^{-1} = \Arm^{-1} \Srm^{-1}(t+\dt) (\Arm^{\dagger})^{-1} \\
&\Srm^{-1}(t+\dt) = \Arm \, \Srm^{-1}(t) \Arm^{\dagger} \; .
\eal

  Again. the metric being invariant under any unitary transformation 
represents a gauge indetermination in the solution of 
Eq.~\eqref{eq:def-exp-connect}, while the definition in 
Eq.~\eqref{eq:basistransf} does not allow for that freedom,
and, therefore, seems a more attractive target.

\section{Evolved bra for constant connection}
\label{app:braevol-constconnect}

  A valid question after Section~\ref{sec:exponential_evolution} 
is whether the evolving bra in Eq.~\eqref{eq:ketevol2} 
corresponds to the instantaneous bra of the evolving ket in 
Eq.~\eqref{eq:ketevol}.
  The instantaneous bra is, turning Eq.~\eqref{eq:ketevol} around,
\beq
\label{eq:braexpmat}
\psi_m^{\phn\mu}(t+\dt) = \psi_m^{\phn\nu}(t) \, 
e^{\dt (\ioh \, H_{\nu}^{\phe\mu} - D_{\nu t}^{\phm\mu})} \; ,
\eeq
which is just the complex conjugate of the coefficients for the ket in 
Eq.~\eqref{eq:ketevol}.
  It is also the solution of the equation of motion given by
\beq
\label{eq:evolbra-alt}
\partial_t \psi_m^{\phn\mu} = \psi_m^{\phn\nu} 
(\ioh \, H_{\nu}^{\phe\mu} - D_{\nu t}^{\phm\mu}) \; ,
\eeq
given that both $H_{\nu}^{\phe\mu}$ and $D_{\nu t}^{\phm\mu}$
are time-independent if $H^{\mu}_{\phe\nu}$ and $D^{\mu}_{\phe\nu t}$
are, as assumed before.
  This equation is derived from $\langle \psi_m | H = -i\hbar \partial_t \langle \psi_m |$,
\bals
&\left (\partial_t \langle \psi_m | \right ) | e^{\mu}\rangle = 
\ioh \langle \psi_m | H | e^{\mu} \rangle \\
&\partial_t \psi_m^{\phn\mu} - \langle \psi_m | \partial_t e^{\mu} \rangle =
\ioh \psi_m^{\phn\nu} H_{\nu}^{\phe\mu} \\
&\partial_t \psi_m^{\phn\mu} - \psi_m^{\phn\nu} D_{\nu\phe t}^{\phe\mu} =
\ioh \psi_m^{\phn\nu} H_{\nu}^{\phe\mu} \; ,
\eals
which, remembering that $D_{\nu\phe t}^{\phe\mu} = - D_{\nu t}^{\phm\mu}$,
is nothing but Eq.~\eqref{eq:evolbra-alt}, or, canonically,
\beqs
-i \hbar \, \covdev_t \psi_m^{\phe\mu} = \psi_m^{\phe\nu} H_{\nu}^{\phe\mu}
\eeqs

   Similarly, Eqs.~\eqref{eq:evoldiffbra} and \eqref{eq:evolbra-alt} 
can be easily shown to be equivalent:
\bals
\partial_t \psi_{m\mu} &= \psi_{m\nu} 
(\ioh \, H^{\nu}_{\phe\mu} + D^{\nu}_{\phe\mu t}) \\
\partial_t ( \psi_m^{\phn\sigma} S_{\sigma\mu}) &= 
\psi_m^{\phn\lambda} S_{\lambda\nu} 
(\ioh \, H^{\nu}_{\phe\mu} + D^{\nu}_{\phe\mu t}) \\
(\partial_t \psi_m^{\phn\sigma}) S_{\sigma\mu}  &= 
\psi_m^{\phn\nu} (\ioh \, H_{\nu\mu} + D_{\nu\mu t}) 
- \psi_m^{\phn\sigma} \partial_t  S_{\sigma\mu}\\
(\partial_t \psi_m^{\phn\sigma}) S_{\sigma\mu}  &= 
\psi_m^{\phn\nu} 
(\ioh \, H_{\nu\mu} + D_{\nu\mu t}-\partial_t  S_{\nu\mu}) \\
\partial_t \psi_m^{\phn\mu} &= \psi_m^{\phn\nu} 
(\ioh \, H_{\nu\sigma} + D_{\nu\sigma t}-\partial_t  S_{\nu\sigma}) 
S^{\sigma\mu} \\
\partial_t \psi_m^{\phn\mu} &= \psi_m^{\phn\nu} 
(\ioh \, H_{\nu}^{\phe\mu} + D_{\nu\sigma t} S^{\sigma\mu} + 
S_{\nu\sigma} \partial_t S^{\sigma\mu}) \\
\partial_t \psi_m^{\phn\mu} &= \psi_m^{\phn\nu} 
(\ioh \, H_{\nu}^{\phe\mu} + D_{\nu\phe t}^{\phe\mu} ) \\
\partial_t \psi_m^{\phn\mu} &= \psi_m^{\phn\nu} 
(\ioh \, H_{\nu}^{\phe\mu} - D_{\nu t}^{\phm\mu}) \quad \mathrm{q.e.d.} 
\eals

  Therefore, the bra coefficients in Eq.~\eqref{eq:braexpmat}, which 
are complex conjugates of the ket coefficients in Eq.~\eqref{eq:ketevol}, 
represent the evolved solution of Eq.~\eqref{eq:evolbra-alt}, which is equivalent
to Eq.~\eqref{eq:evoldiffbra}, that gives the evolved solution of
Eq.~\eqref{eq:ketevol2}. 
  It all closes the consistency circle showing that $\psi_{m\mu} (t+\dt)$
of Eq.~\eqref{eq:ketevol2} and $\psi_m^{\phn\mu}(t+\dt)$ of 
Eq.~\eqref{eq:braexpmat} are related by
\beq
\label{eq:braequiv}
\psi_{m\mu} (t+\dt) = \psi_m^{\phn\nu}(t+\dt) \, S_{\nu\mu}(t+\dt) \; ,
\eeq
as they should.




\end{document}